\documentclass[aps,prl,twocolumn,superscriptaddress]{revtex4-2}
\usepackage{amsmath, amssymb,  graphicx,bbm}

\usepackage[colorlinks=true ,urlcolor=blue,urlbordercolor={0 1 1}]{hyperref}

\usepackage{color}
\usepackage{xcolor}
\usepackage{braket}
\usepackage{natbib}
\usepackage[utf8]{inputenc}
\usepackage{bm}
\usepackage{verbatim}
\usepackage{graphicx}
\usepackage{amsmath}
\usepackage{color}
\usepackage{float}
\usepackage{bbm}
\usepackage{amssymb}
\usepackage{slashed}
\usepackage{wasysym,bm,bbm,dsfont,braket}

\begin{document}

\title{Spin-charge separation in the triangular-lattice Hofstadter-Hubbard model}

\author{Yuntian Gu}
\thanks{These authors contributed equally to this work.}
\affiliation{School of  Intelligence Science and Technology, Peking University, Beijing 100871, China.}

\author{Hui Yang}
\thanks{These authors contributed equally to this work.}
\affiliation{Department of Physics and Astronomy, University of Pittsburgh, PA 15213, USA}

\author{Zhehao Dai}
\email{daizhehao@pitt.edu}
\affiliation{Department of Physics and Astronomy, University of Pittsburgh, PA 15213, USA}

\author{Yantao Wu}
\email{yantaow@iphy.ac.cn}
\affiliation{Institute of Physics,
Chinese Academy of Sciences, Beijing 100190, China.}

\begin{abstract}
Recent experiments in moiré materials have enabled the realization of
a variety of exotic quantum phases. 
In this context, the Hofstadter-Hubbard model has been proposed as a possible setting for hosting chiral spin liquid.
Concurrently, significant progress has been recently made in the computational methods for two-dimensional many-body fermion systems, which makes numerically studying this challenging model a real possibility in genuine 2D geometry.
Motivated by these advances, we investigate the putative chiral spin liquid phase in the triangular-lattice Hofstadter-Hubbard model using variational Monte Carlo with neural quantum states (NQS) and projected entangled pair states (PEPS). 
We observe spin-charge separation directly in real space through numerical spin-pumping simulation and real-time spin and charge motion. 
In addition, in the context of anyonic superconductivity conjectured in this model, we find a positive two-electron binding energy on small systems, but it decreases below our numerical resolution as the system size increases.
Our work demonstrates NQS and PEPS as powerful tools, capable of cross-checking each other, for diagnosing topological order and fractionalized excitations in strongly correlated electronic systems.
\end{abstract}
\maketitle

\textit{Introduction.} 
Spin-charge separation, in which an electron's spin and charge quantum numbers are carried by distinct excitations, is one of the most striking examples of fractionalization.
It is well established in one-dimensional repulsive electron systems described by Luttinger-liquid theory~\cite{haldane1981luttinger,voit1995one,giamarchi2003quantum}.
In two dimensions, theoretical routes to electron fractionalization have been developed within resonating-valence-bond, parton field theory, and topological-order frameworks~\cite{anderson1973resonating,kalmeyer1987equivalence,wen1990topological,senthil2000z,lee2006doping}.
Such fractionalization is expected in spin liquids, for which several promising material candidates exist, although no definitive observation of spin-charge separation has been established~\cite{balents2010spin}.
Numerical studies have investigated candidate spin-liquid regimes in both spin models and electronic Hubbard models using complementary diagnostics, including entanglement spectra, spin pumping, scalar chirality, and the absence of conventional magnetic order~\cite{bauer2014chiral,gong2014emergent,shirakawa2017ground,szasz2020chiral,Yan2011Kagome,Depenbrock2012Kagome,Jiang2012Entanglement,Iqbal2013Kagome,Liao2017Kagome,He2017DiracKagome,Lauchli2019Kagome,Kaneko2014Triangular,Zhu2015Triangular,Hu2015Triangular,Saadatmand2016Triangular,Hu2019DiracTriangular,He2014CSL,Gong2014CSL,Bauer2014CSL,Gong2015Kagome,Zhu2015KagomeCSL,Wietek2015Kagome,Wietek2017TriangularCSL,Hickey2017Melting,Cookmeyer2021CSL,Zhang2024SquareCSL,Hickey2016HaldaneHubbard,Szasz2020HubbardCSL,Chen2022HubbardCSL,Zhu2024HubbardCSL}.
However, direct real-space observation of spin-charge separation remains challenging because reliable numerical methods for correlated fermions in genuine two-dimensional geometries have only recently become available.

In this letter, we present direct real-space evidence of spin-charge separation on finite two-dimensional lattices using variational Monte Carlo (VMC) algorithms on neural quantum states (NQS)~\cite{carleo2017solving,gu2026solving,gu2026pareto} and projected entangled pair states (PEPS)~\cite{Wu_2026,wu2025realtime}.
The model we study is the triangular-lattice Hofstadter-Hubbard model~\cite{kuhlenkamp2024chiral,divic2026chiral,divic2025anyon,gallegos2026quantum,niu2025thermodynamic,kuhlenkamp2025robust}, which recently attracted great interest for its relevance to moir\'e materials and and its potential to realize chiral spin liquid (CSL)~\cite{kalmeyer1987equivalence,wen1990topological} and unconventional superconductivity. The Hamiltonian consists of hopping $te^{i\theta_{ij}}$ and Hubbard interaction $U$.
The phase $\theta_{ij}$ is chosen so that each elementary triangle has flux $\pi/2$.
The large lattice constants of moir\'e superlattices make substantial magnetic flux per unit cell accessible 
with realistic magnetic fields
~\cite{dean2013hofstadter,hunt2013massive}.
At half filling (one electron per site), density matrix renormalization group (DMRG) calculations suggest a CSL at intermediate $U$~\cite{kuhlenkamp2024chiral,divic2026chiral,gallegos2026quantum}.
In this phase, an electron can fractionalize into a neutral spin-$1/2$ spinon and a spinless charge-$e$ chargon~\cite{lee2006doping,gallegos2026spinless}.

\textit{Result summary.}
We give a summary of our findings here.
\begin{enumerate}
\item
With open boundary conditions, finite-size scaling from $U=7$ to $U=13$ shows the opening of a single-electron gap while the edge spin gap remains closed.
\item Quantized spin response is observed at $U=13$ upon spin-flux insertion, whereas the response to charge-flux insertion vanishes.
In particular, our pumped spin depends linearly on the spin flux, whereas spin response generally oscillates around a straight line in quasi-1D geometries.
\item
The spin and charge degrees of freedom separate clearly in real-time dynamics.
After an electron insertion on the edge, the charge propagates quickly into the bulk while the spin slowly moves around the edge.
\item A positive pair-binding energy is observed for small system sizes, consistent with previous studies~\cite{divic2025anyon}, but it decreases below our numerical resolution with increasing system size.
\end{enumerate}

\begin{figure}[h]
    \centering
\includegraphics[width=0.8\linewidth]{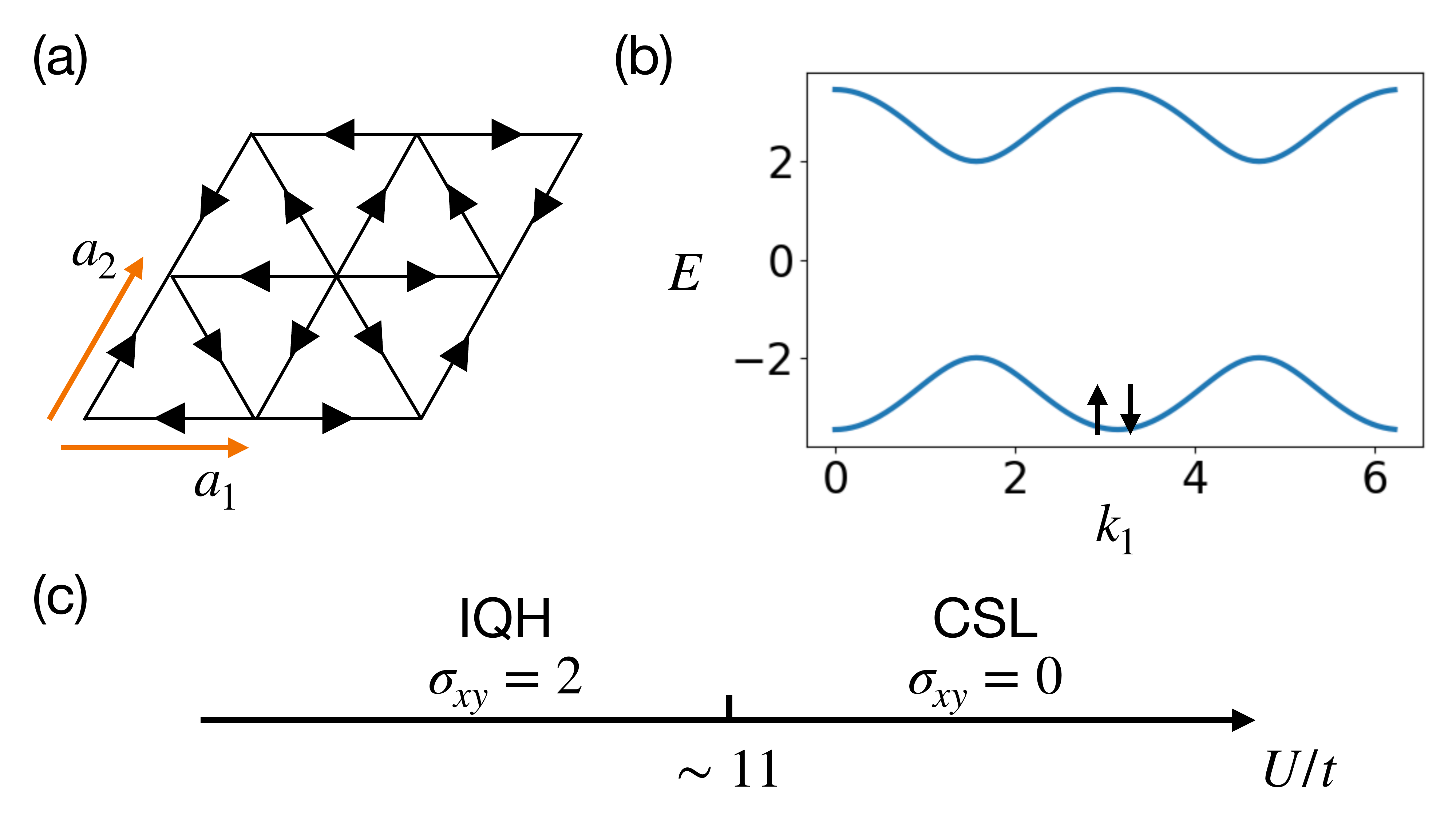}
\caption{Triangular-lattice Hofstadter-Hubbard model.
(a) Gauge choice for the nearest-neighbor hopping shown in a $2\times 2$ unit cell, 
${\bf a}_1=(1,0)$, and ${\bf a}_2=(1/2,\sqrt{3}/2)$.
An arrow from $j$ to $i$ denotes a hopping phase $e^{i\theta_{ij}}=i$; the reverse hopping is fixed by Hermiticity.
The phases give flux $\Phi=\pi/2$ through each triangular plaquette when traversed clockwise.
The gauge is chosen so that the sixfold rotation about the center acts as $C_6 c_{\bf r} C_6^\dagger=c_{C_6({\bf r})}$.
(b) Single-particle dispersion along $k_1$, at $k_2=0$, for the magnetic unit cell spanned by ${\bf a}_1$ and $2{\bf a}_2$, where $e^{ik_1}$ is the eigenvalue of the magnetic translation $T_{{\bf a}_1}$.
At half filling, the lower band is filled by both spins giving a total Chern number $C=C_\uparrow+C_\downarrow=2$.
(c) Schematic phase diagram with respect to $U/t$, showing the weak-coupling IQH phase and the intermediate-coupling CSL.}
\label{fig:model_band_phase_diagram}
\end{figure}

\textit{Model.}
The triangular-lattice Hofstadter-Hubbard model is
\begin{align}
H\hspace{-3pt}=-\hspace{-5pt}\sum_{\langle ij\rangle,\sigma}\hspace{-3pt}\left(te^{i\theta_{ij}}c^\dagger_{i\sigma}c_{j\sigma}+\mathrm{h.c.}\right)\hspace{-3pt}+U\hspace{-2pt}\sum_i(n_{i\uparrow}-\frac{1}{2})(n_{i\downarrow}-\frac{1}{2})
\label{eq:model}
\end{align}
Here, $c_{i\sigma}^{\dagger}$ ($c_{i\sigma}$) creates (annihilates) an electron with spin $\sigma=\uparrow,\downarrow$ at site $i$, $n_{i\sigma}=c^\dagger_{i\sigma}c_{i\sigma}$, and $n_i=\sum_\sigma n_{i\sigma}$. 
$U$ is the onsite Hubbard interaction and $t$ is the nearest-neighbor hopping amplitude. The phase $\theta_{ij}$ implements a magnetic flux of $\pi/2$ through each triangle, as illustrated in Fig.~\ref{fig:model_band_phase_diagram}(a). In the following, we take $t=1$ as the energy unit.

The Hamiltonian preserves global $U(1)_c$ charge conservation and $SU(2)$ spin-rotation symmetry, while the magnetic flux explicitly breaks time-reversal symmetry. As shown in Fig.~\ref{fig:model_band_phase_diagram}(b), the lower and the upper band carry Chern numbers $C=1$ and $C=-1$ per spin, respectively. At half filling with one electron per site, the spin-up and spin-down electrons completely occupy the lower band with a total Chern number $C=C_\uparrow+C_\downarrow=2$; thus, the noninteracting ground state realizes an integer quantum Hall (IQH) phase with a quantized charge response $\sigma_{xy}=2\frac{e^2}{h}$ and a quantized spin Hall response $\sigma_{xy}^s=2\frac{\hbar}{8\pi}$.

Previous numerical studies suggest that the onsite repulsion suppresses charge fluctuations and drives the system from the IQH phase into a CSL near $U_c\approx 11$ (Fig.~\ref{fig:model_band_phase_diagram}(c))~\cite{kuhlenkamp2024chiral,kuhlenkamp2025robust,divic2026chiral,gallegos2026quantum,chen2026topological,divic2025anyon}.
The CSL has zero charge Hall response, $\sigma_{xy}=0$, while its spin Hall response remains quantized at $\sigma_{xy}^s=2\frac{\hbar}{8\pi}$. 
Theory predicts that the CSL is topologically equivalent to a bosonic $\nu=1/2$ Laughlin state in the spin sector, with a chiral edge mode for spins~\cite{kalmeyer1987equivalence,kuhlenkamp2024chiral,divic2026chiral,gallegos2026quantum}.
At still stronger interaction, the model is expected to enter a $120^\circ$ antiferromagnetic phase~\cite{kuhlenkamp2024chiral,divic2026chiral}.

\begin{figure}[h]
    \centering
\includegraphics[width=1.0\linewidth]{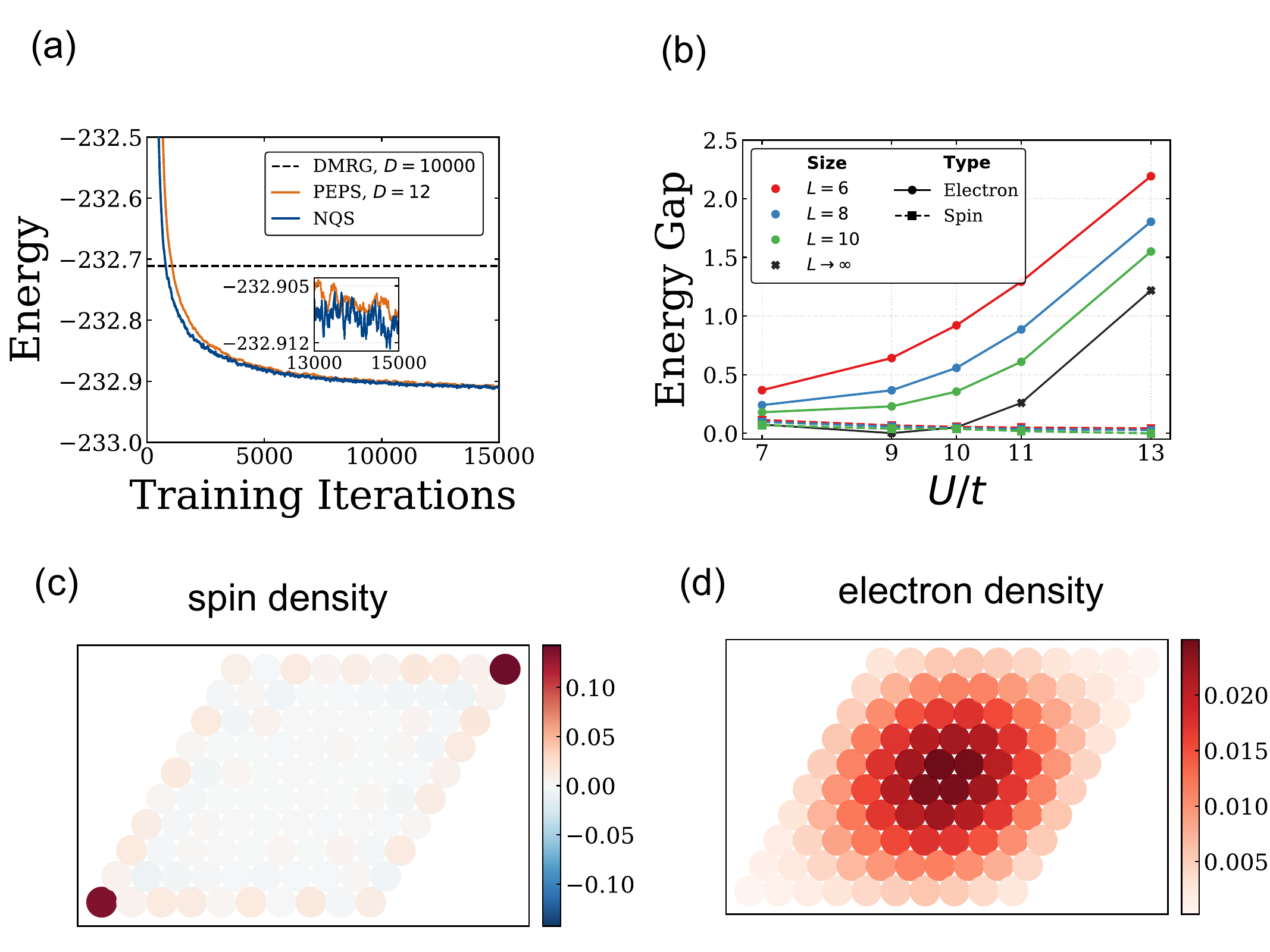}
\caption{Ground state and excited states with open boundary conditions.
(a) Ground-state energy comparison among DMRG, PEPS, and NQS on an $8\times 8$ system at $U=13$.
(b) Single-electron gap $E_0(N+1, S_z=1/2)-E_0(N, S_z=0)$ (solid lines) and spin gap $E_0(N,S_z=1)-E_0(N,S_z=0)$ (dashed lines) as functions of interaction strength $U$ for $L=6$ (green), $L=8$ (blue), and $L=10$ (red).
The grey line represents the electron gap extrapolated to $L\rightarrow\infty$.
(c,d) Spin density and excess charge density of the lowest-energy state with one electron above half-filling at $U=13$.
The spin density is concentrated near the boundary, whereas the excess charge density is mainly distributed in the bulk, giving a real-space diagnostic of spin-charge separation in the CSL.}
\label{fig:ground_state}
\end{figure}

\textit{Ground and excited states.}
We first benchmark our numerical methods by comparing ground-state energies obtained with DMRG, PEPS, and NQS. The DMRG calculation uses a bond dimension up to $D=10000$ and a truncation error below $10^{-4}$, while the PEPS calculation uses bond dimension $D=12$. For NQS, we use the recently introduced Accurate Convolutional ansatz for lattice Electrons (ACE)~\cite{gu2026pareto}.
As shown in Fig.~\ref{fig:ground_state}(a), PEPS and NQS yield similar variational energies, both below the DMRG result. The DMRG calculation required $143.7$ hours of wall time on a CPU cluster to reach a bond dimension $D=10000$ with $U(1)$ charge-conservation and $U(1)$ $S_z$-conservation, while PEPS and NQS required approximately 196 and 15 hours on 8 A100 GPUs, respectively. Within this benchmark, NQS yields the lowest variational energy and requires 
the least wall-time; we therefore use NQS as the main method for static ground-state calculations.

To characterize the ground-state phase transition, we calculate the single-electron gap $E_0(N+1, S_z=1/2)-E_0(N, S_z=0)$ and the spin gap $E_0(N,S_z=1)-E_0(N,S_z=0)$ with open boundary conditions for several system sizes at $U=7,9,10,11,13$. The chemical potential vanishes because of particle-hole symmetry.
Here, $N = L^2$ is the total electron number at half filling and $S_z$ is the $z$ component of the total spin.
As shown in Fig.~\ref{fig:ground_state}(b), we find that for $U\le 10$, both the electron gap and the spin gap decrease toward zero with increasing system size, consistent with the chiral edge mode of the IQH phase.
For $U\ge11$, the electron gap extrapolates to a nonzero value, while the spin gap remains close to zero, consistent with a transition from the IQH phase to the CSL.

To further investigate the properties of the CSL, we calculate the spatial distributions of the spin and charge densities for the ground state with an additional electron relative to half filling at $U=13$. As shown in Fig.~\ref{fig:ground_state}(c), the spin density is strongly localized on the edge and peaks at two corners.
The edge localization is consistent with a gapless spin mode, while the additional accumulation at the corners reflects their reduced velocity near the corners.
In contrast, the charge density of the $1e$ excitation is predominantly in the bulk (Fig.~\ref{fig:ground_state}(d)), reflecting the absence of a gapless charge mode at the boundary in the CSL. 
The contrasting spatial distributions of the spin and charge densities provide direct real-space evidence of spin-charge separation on a two-dimensional lattice.

\begin{figure}[h]
    \centering
\includegraphics[width=1.0\linewidth]{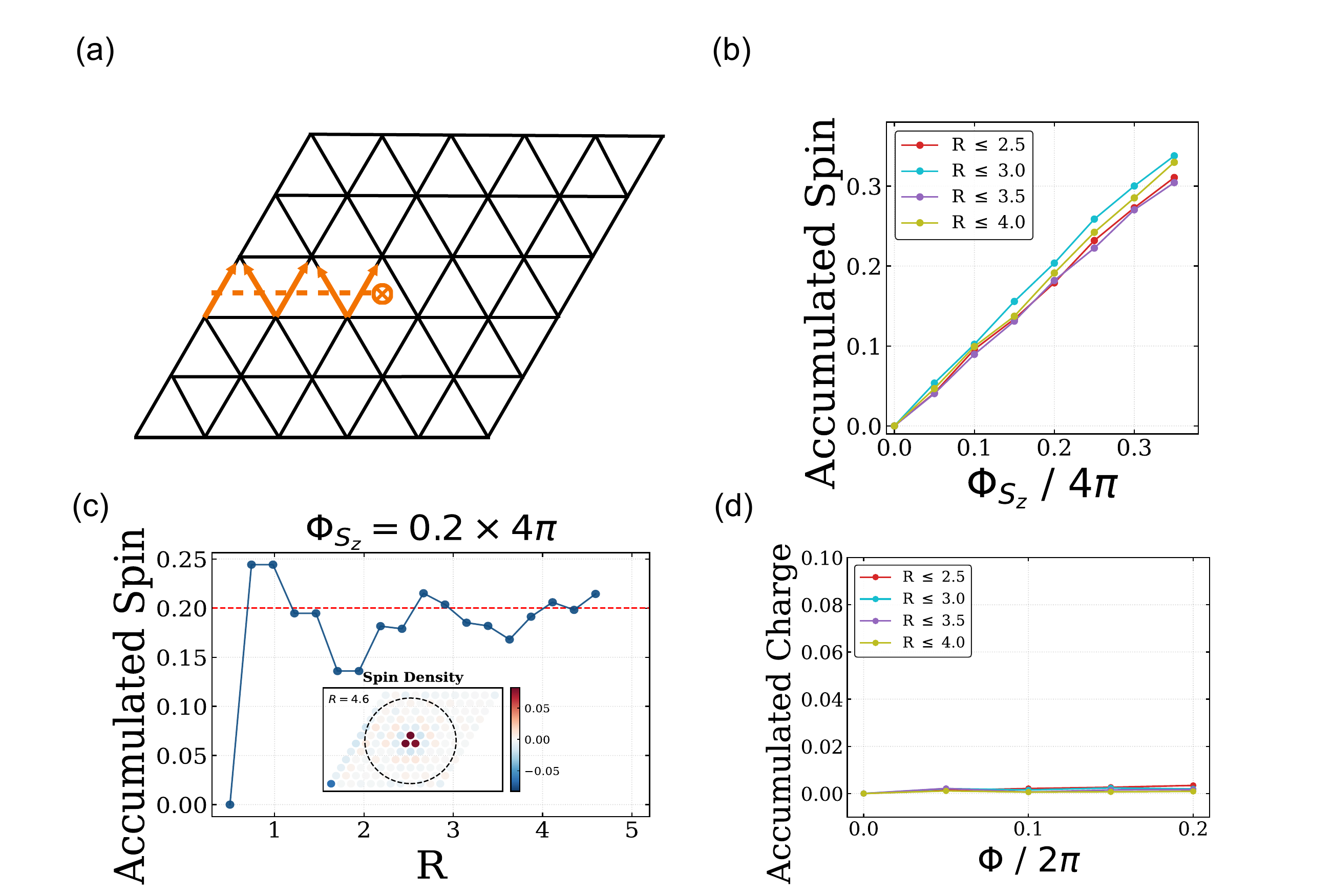}
\caption{Spin and charge pumping in the CSL phase at $U=13$ with open boundary conditions.
(a) Illustration of flux insertion.
The flux is placed at the orange cross, and hoppings that cross the dashed branch cut acquire $\theta_{ij}\rightarrow \theta_{ij} + \sigma_z\Phi_{S_z}/2$ for spin-flux insertion or $\theta_{ij}\rightarrow \theta_{ij}+\Phi$ for charge-flux insertion.
(b) Spin $S_z(R)$ enclosed within radius $R$ of the flux as a function of spin flux $\Phi_{S_z}$ on a $12\times12$ lattice.
For $2\le R\le 4$, the linear response approaches the quantized spin Hall conductance $\sigma^s_{xy}$.
(c) Radial profile of the accumulated spin at fixed spin flux $\Phi_{S_z}=0.2\times4\pi$ on a $12\times12$ cluster.
The spin approaches the theoretical prediction before the circle reaches the boundary; the inset shows the corresponding local $S_z$ distribution. 
(d) Accumulated charge under charge-flux insertion on a $10\times10$ lattice.
The charge response remains negligible over the same flux range, distinguishing the CSL from the IQH state.}
\label{fig:spin_charge_pumping}
\end{figure}

\textit{Spin pumping.} 
To test the theoretically predicted topological spin transport, we study the response to the adiabatic insertion of a spin-dependent flux.
In analogy with bosonic Laughlin states, locally inserting an $S_z$ flux $\Phi_{S_z}$ in a CSL binds an accumulated spin $S_z=\sigma^{s}_{xy}\Phi_{S_z}$. This response is topologically quantized: a $4\pi$ spin flux binds spin $\hbar$~\cite{laughlin1981quantized,kalmeyer1987equivalence,gong2014emergent,bauer2014chiral,szasz2020chiral,gallegos2026quantum}.
We choose the convention that a spin flux $\Phi_{S_z}$ corresponds to inserting the flux $\Phi_{S_z}/2$ for up spins and $-\Phi_{S_z}/2$ for down spins.
Note that the IQH phase produces the same spin response because the $2\pi$ ($-2\pi$) flux inserted for spin-up (spin-down) electrons attracts $1$ ($-1$) electron; the two spin species together contribute spin $\hbar$.
However, the IQH phase also has a nonzero charge response when we insert the same flux for spin up and spin down.
Thus, the quantized spin Hall response in the absence of charge response is the hallmark of the CSL topological order.
 
To test this topological response numerically, we insert a spin flux $\Phi_{S_z}$ at the center of the system and compute the spin accumulated in a circle with radius $R$ around the flux.
As shown in Fig.~\ref{fig:spin_charge_pumping}(a), we implement the spin flux $\Phi_{S_z}$ (marked by an orange ``$\otimes$'') by changing the hoppings along the orange arrows according to $\theta_{ij}\rightarrow \theta_{ij} + \sigma_z\Phi_{S_z}/2$.
We measure the total spin enclosed within a radius $R$ of the flux-insertion point ${\bf{r}_0}$: $S_z(R)=\sum_{|{\bf{r}} - {\bf{r}_0}|<R}S_{\bf{r}}^z$.
The flux is expected to pump spin $S_z=\sigma^{s}_{xy}\Phi_{S_z}$ from the edge into a region within a few correlation lengths $\xi$ of the center. Thus, when $R\gg \xi$ and $\sqrt{3}L/4-R\gg \xi$, we expect $S_z(R)=\sigma^{s}_{xy}\Phi_{S_z}$.
As shown in Fig.~\ref{fig:spin_charge_pumping}(b), the slope of $S_z(R)$ approaches this quantized value for $2\le R\le 4$ with the system size $L=12$.
Small deviations from quantization are finite-size effects.
Fig.~\ref{fig:spin_charge_pumping}(c) shows $S_z(R)$ as a function of $R$ for a fixed flux $\Phi_{S_z}=0.2\times 4\pi$.
The integrated spin oscillates and converges to the theoretical prediction $0.2\hbar$ before the circle reaches the boundary.
The inset shows the spin density on each site, with positive spin concentrated near the center.
In contrast, no charge pumping is observed at $U=13$ when we insert the same flux $\Phi$ for both spins [Fig.~\ref{fig:spin_charge_pumping}(d)]. Together, the quantized spin response and vanishing charge response diagnose the topological response of the CSL.

\begin{figure}[h]
    \centering
\includegraphics[width=1.0\linewidth]{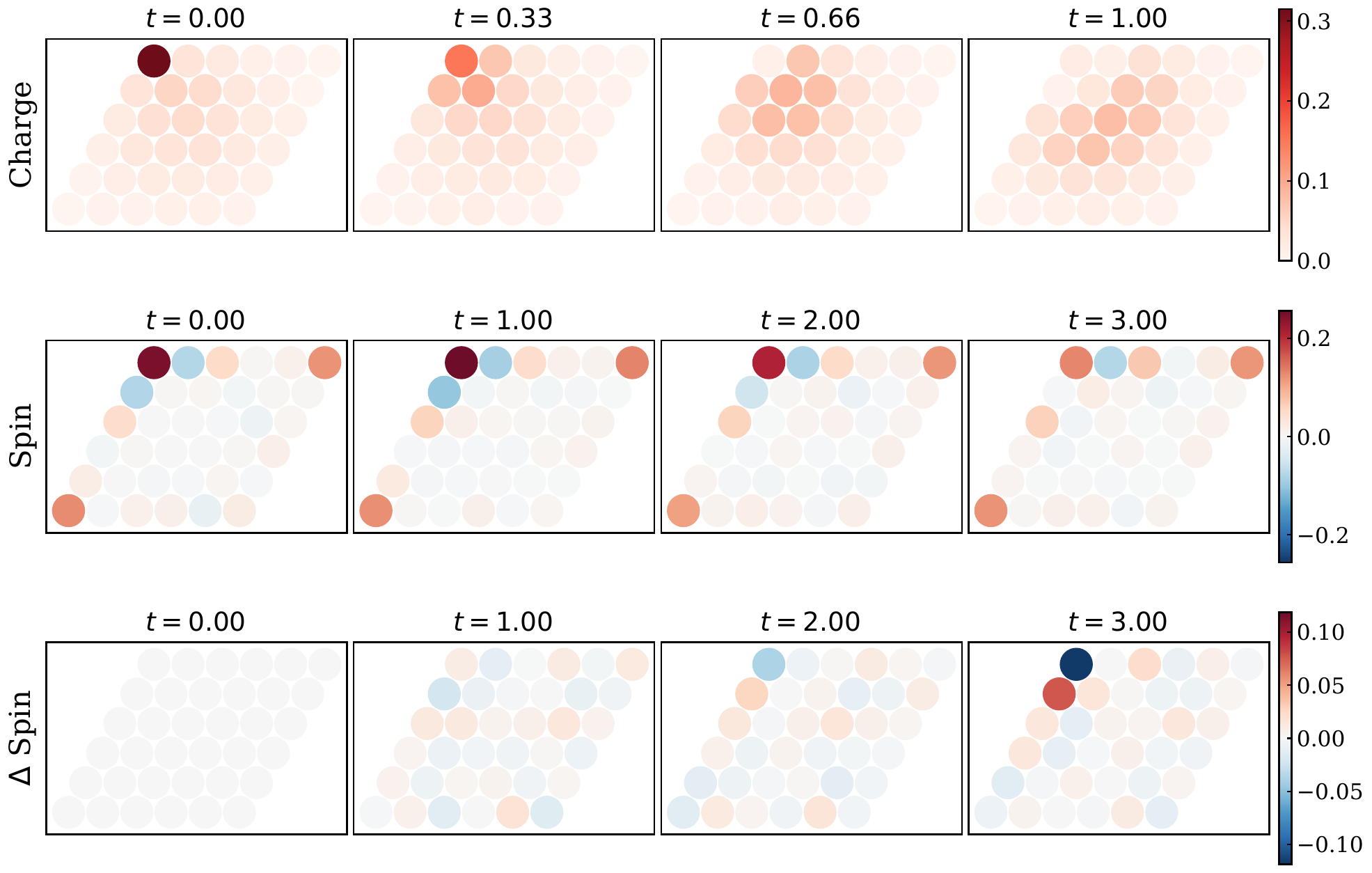}
\caption{Real-time PEPS dynamics of an injected electron in the CSL phase at $U=13$ on a $6\times6$ cluster.
The initial state contains one additional electron pinned near the upper-left corner by spin-dependent local attractive potentials $h_{\rm pin}^{\uparrow}=4.4$ and $h_{\rm pin}^{\downarrow}=3.6$.
After the potential is removed at $t=0$, the excess charge rapidly spreads into the bulk, while the spin density remains near the boundary and propagates along the edge.
The contrasting charge and spin motion provides a direct dynamical signature of spin-charge separation.}
\label{fig:dynamics}
\end{figure}

\textit{Dynamics.}
We simulate real-time dynamics using the time-dependent VMC algorithm for PEPS~\cite{wu2025realtime}.
We first create a localized electron near the corner by applying spin-dependent local attractive potentials at one of the obtuse corners, with $h_{\rm pin}^{\uparrow}=4.4$ and $h_{\rm pin}^{\downarrow}=3.6$.
After preparing this initial state, we remove the pinning potential, evolve the state under the original Hamiltonian in Eq.~\ref{eq:model}, and monitor the local spin and charge densities at each time step.
We find that the spin and charge exhibit markedly different behavior during the time evolution.
As shown in Fig.~\ref{fig:dynamics}, the spin density remains strongly localized near the edge and propagates slowly along the boundary, reflecting the chiral edge dynamics of the CSL phase.
In contrast, the charge density quickly propagates into the bulk and spreads over the system, approaching an almost uniform distribution at later times. 
This result provides a direct dynamical signature of spin-charge separation: the injected electron at the corner splits into spin and charge excitations with markedly different dynamics.

\textit{Pairing.}
Having established spin-charge separation in the CSL, we now turn to pairing upon doping. Doping a quantum spin liquid has long been proposed as a route to superconductivity~\cite{laughlin1988superconducting,Witten_anyonsc,wen_anyon_1991,wen_anyon_2013,jiang2021superconductivity,song_dopeCSL,jiang_dopeCSL,han_anyon_sf}. For the present model, recent studies have proposed enhanced electron pairing near the IQH-CSL transition and reported chiral superconductivity at finite carrier density in quasi-1D cylinders with widths up to 6~\cite{divic2025anyon,kuhlenkamp2025robust,chen2026topological,niu2025thermodynamic}. Here we instead probe the dilute two-electron limit on a 2D system symmetric under 6-fold rotations.
With periodic boundary condition, by computing the excitation energies of the $1e$ state, $E_{1e}=E_0(N+1, S_z=1/2)-E_0(N, S_z=0)$, and the $2e$ state, $E_{2e}=E_0(N+2, S_z=0)-E_0(N, S_z=0)$, we obtain the pair-binding energy $E_b=2E_{1e}-E_{2e}$, for which $E_b>0$ indicates binding.

\begin{figure}[h]
    \centering
\includegraphics[width=1.0\linewidth]{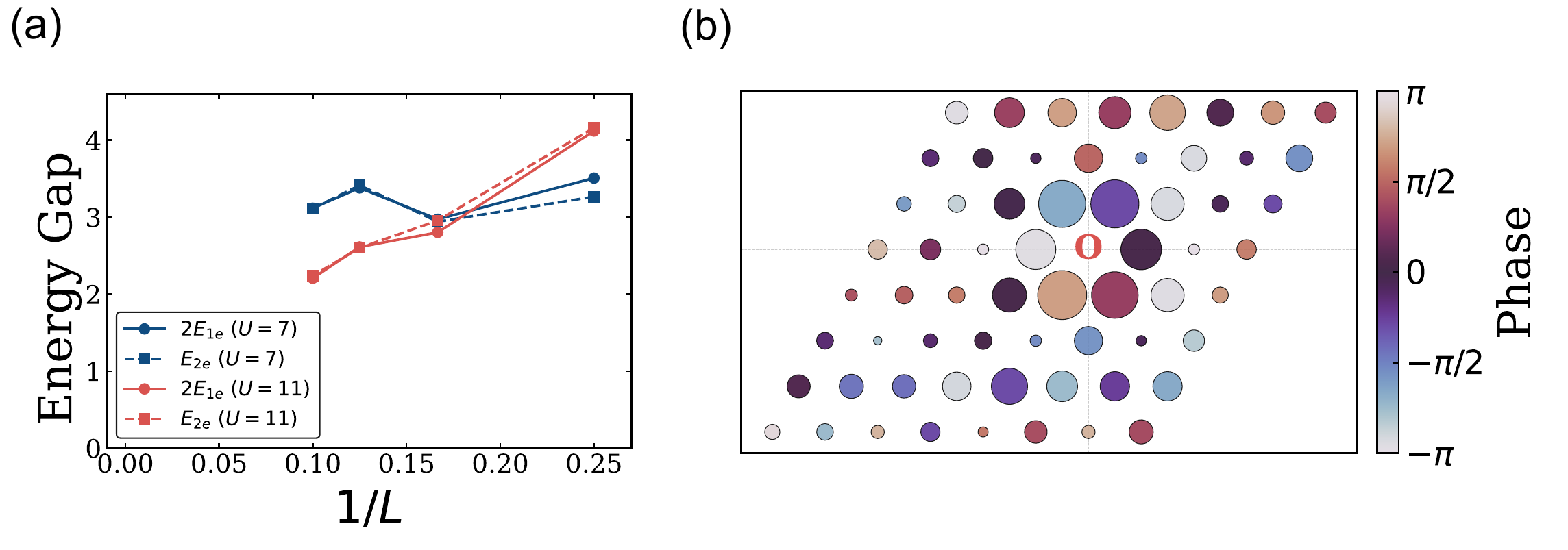}
\caption{Pair energy, single electron energy, and pair wave function near the IQH-CSL transition with periodic boundary conditions.
(a) Size dependence of two times the charge-$1e$ energy $2E_{1e}=2(E_0(N+1, S_z=1/2)-E_0(N, S_z=0))$ and charge-$2e$ energy $E_{2e}=E_0(N+2, S_z=0)-E_0(N, S_z=0)$.
At $U=7$, the positive binding energy observed on small clusters decreases with increasing system size and falls below numerical resolution on the largest clusters.
Near the critical interaction, $U=11$, the binding energy is consistent with zero for all sizes studied.
(b) Spatial structure of the spin-singlet pair wave function $\Delta({\bf r})=\bra{\Psi_{2e}}(c^\dagger_{{\bf r}\uparrow}c^\dagger_{0\downarrow}-c^\dagger_{{\bf r}\downarrow}c^\dagger_{0\uparrow})\ket{\Psi_{\rm GS}}$ for $U=11$ and $L=8$, showing the phase winding of the lowest charge-$2e$ state.}
\label{fig:pairing}
\end{figure}

As shown in Fig.~\ref{fig:pairing}(a), in the IQH phase at $U=7$, $E_b$ is positive at $L=4$, consistent with previous finite-cylinder and exact-diagonalization results~\cite{divic2025anyon}, but it decreases with increasing system size. 
For $L=10$, the binding energy is below our numerical resolution.
Moreover, near the critical point, $E_b$ is close to zero or negative for all system sizes in our simulations, as shown in Fig.~\ref{fig:pairing}(a). Our results therefore constrain two-electron binding in the dilute limit but do not directly determine superconductivity at finite carrier density. Within the present resolution, whether the single-electron gap in the bulk closes near the phase transition also remains unresolved.

We next analyze the symmetry and pair wave function of the lowest charge-$2e$ state. For $U=11$ and $L=8$, the optimized state $\ket{\Psi_{2e}}$ has a dominant component in the $C_6$ angular-momentum sector $\ell=-1$ (mod 6), consistent with a previous exact-diagonalization result for a $4\times4$ system~\cite{divic2025anyon}. We calculate the pair wave function $\Delta({\bf{r}})=\bra{\Psi_{2e}}(c^\dagger_{{\bf{r}}\uparrow}c^\dagger_{0\downarrow}-c^\dagger_{{\bf{r}}\downarrow}c^\dagger_{0\uparrow})\ket{\Psi_{\rm GS}}$. As shown in Fig.~\ref{fig:pairing}(b), the pair wave function exhibits a $2\pi$ winding around the origin.
Previous work predicts that, in the gauge shown in Fig.~\ref{fig:model_band_phase_diagram}(a), $\Delta({\bf{r}})$ acquires a phase $e^{i2\pi/6}$ under a $C_6$ rotation~\cite{divic2025anyon}.
Our result approximately respects this symmetry but retains appreciable weight in other symmetry sectors.
Such mixing is consistent with several nearly degenerate charge-$2e$ states and the absence of pairing gap; the phase texture alone therefore does not establish a stable bound pair.

\textit{Conclusion.}
In summary, we study the half-filled triangular lattice Hofstadter-Hubbard model using recently developed NQS and PEPS algorithms. 
We confirm the transition from IQH to CSL by measuring the electron gap and the spin gap  on the edge as a function of $U$.
In the CSL phase, we numerically observe spin-charge separation via PEPS simulation of real-time evolution and NQS simulation of excited states and spin pumping.
We further examine the pairing tendency upon doping.
We measure the charge 2e gap and charge 1e gap at $U=7t$ and $U=11t$ under periodic boundary conditions for system sizes up to $10\times 10$.
We observe a large pairing gap at $U=7t$ for the $4\times4$ system, consistent with previous results~\cite{divic2025anyon}. However, no pairing gap is observed for other larger system sizes.
The situation deserves further study in the thermodynamic limit, especially near the critical point.

More broadly, the complementary use of NQS and PEPS allows us to access both ground-state topological properties and real-time fractionalized dynamics.
We expect these numerical methods to continually develop, enabling further study of other strongly-correlated systems.

\textit{Note added.} Ref.~\cite{roth2026large}, which appeared on the same day, also study the triangular-lattice Hofstadter-Hubbard model using NQS. We thank the authors for coordinating submissions of our papers. 

\textit{Acknowledgments.}
The NQS calculations use the LaQX codebase~\cite{bytedance2026laqx}. Z.D. and H.Y. are supported by the startup funds at the University of Pittsburgh.
Y.W. acknowledges support from the Chinese Academy of Sciences (CAS) under Grant No. YSBR-150 and a start-up grant from IOP-CAS. 
The simulations were supported by the IOP-CAS computing facilities and the center for research computing and data at the University of Pittsburgh.
The data that support the findings of this article are openly available \cite{data}; embargo periods may apply.

\bibliographystyle{apsrev4-2}
\bibliography{draft}

\end{document}